\hfuzz 5pt
\raggedbottom
\def\pmb#1{\setbox0=\hbox{#1}
  \kern-.025em\copy0\kern-\wd0
  \kern.05em\copy0\kern-\wd0
  \kern-.025em\raise.0433em\box0 }
\font\bigrm=cmbx12
\font\smrm=cmr8
\magnification=\magstep1
\openup 2\jot
\def\bn{{\pmb{$\nabla$}}}
\def\v{{\pmb{$v$}}}
\def\z{{\pmb{$z$}}}
\def\x{{\pmb{$x$}}}
\def\pa{\partial}
\def\H{{\cal H}}
\def\M{{\cal M}}
\def\C{{\cal C}}
\def\J{{\cal J}}
\def\A{{\cal A}}
\def\P{{\cal P}}
\def\E{{\cal E}}
\def\e{{\rm e}}
\def\sech{{\rm sech}}
\vbox{\vskip1.0truein}
\centerline{\bigrm A Hamiltonian weak-wave model for
shallow-water flow}
\bigskip
\centerline{C{\smrm AROLINE} N{\smrm ORE}}
\smallskip
\centerline{\sl Laboratoire de Physique Statistique, Ecole
Normale Sup\'erieure,}
\vskip-0.1truein
\centerline{\sl 24, rue Lhomond, 75231 Paris 05 Cedex, France}
\smallskip
\centerline{and}
\medskip
\centerline{T{\smrm HEODORE} G. S{\smrm HEPHERD}}
\smallskip
\centerline{\sl Department of Physics, University of Toronto,
Toronto M5S 1A7 Canada}
\vskip1.0truein
\centerline{ABSTRACT}
\bigskip
A reduced dynamical model is derived which describes the
interaction of weak inertia-gravity waves with nonlinear vortical
motion in the context of rotating shallow-water flow. The formal
scaling assumptions are (i) that there is a separation in
timescales between the vortical motion and the inertia-gravity
waves, and (ii) that the divergence is weak compared to the
vorticity. The model is Hamiltonian, and possesses conservation laws
analogous to those in the shallow-water equations. Unlike the
shallow-water equations, the energy invariant is
quadratic. Nonlinear stability theorems are derived for this
system, and its linear eigenvalue properties are investigated in
the context of some simple basic flows.

\vfill\eject
\bigskip\noindent
{\bf 1. Introduction}
\bigskip
An important area of research in geophysical fluid dynamics is
the development of simplified models. Simplified models are
reduced models of the full hydrodynamical equations of motion
that retain the essential physics of the problem under
investigation, yet are reduced in complexity and are
therefore more amenable to theoretical analysis.

A generic property of atmospheric and oceanic dynamics is the
existence of nonlinear vortical motion, evolving on an advective
timescale, together with relatively fast acoustic and gravity
wave motion. While both kinds of motion can be
identified observationally, it is generally the case that the
vortical motion overwhelmingly dominates the fast wave motion for
both meso-scale and large-scale dynamics. This fact has led to a
class of simplified models known as balanced models, which filter
the fast wave motion and describe only the nonlinear vortical
motion. Examples of such models include the barotropic vorticity
equation, the quasi-geostrophic equations, and the
semi-geostrophic equations (see e.g. Pedlosky 1987).

Yet while the fast wave motion is observationally found to be
relatively weak, it is nevertheless ubiquitous. Moreover,
there are strong theoretical reasons for believing that
balanced models cannot be exact (even in principle), and that
some degree of fast motion is inevitable. There is,
therefore, considerable interest in the problem of the
interaction between (slow) nonlinear vortical motion and (fast) weak
waves. Clearly it would be desirable to avoid dealing with
the full primitive equations of motion in theoretical
investigations of this nature. Yet simplified
models describing this kind of interaction do not exist, except for
very special cases.

In this paper we address this problem of constructing
simplified models describing the interaction of nonlinear
vortical motion with weak waves under a separation in
timescales. A traditional approach is
through perturbation methods, employing an asymptotic expansion with
the timescale ratio between the wave and vortical motion (i.e.
the Mach, Froude, or Rossby number) serving as
the small expansion parameter. However, such methods are generally
carried only to leading order. This is, perhaps, not surprising
in light of the fact that the higher-order corrections in
traditional expansions will generically exhibit unbounded secular
growth in time (Warn {\it et al.} 1995). Yet to obtain
non-trivial coupling between the wave and vortical motion, one has
to go beyond leading order.

A rather different approach is to argue that one should maintain
the fundamental dynamical properties of the original system,
notably its conservation properties. The importance of
maintaining conservation properties in dynamical simplifications
has been long recognized (Lorenz 1960, Sadourny 1975). More
recently, it has been realized (e.g. Salmon 1983) that the most
efficient way to do this is through maintaining the Hamiltonian
structure of the dynamics: the simplifications in question
invariably affect the conservative part of the dynamics, which
for nearly every model in geophysical fluid dynamics is
Hamiltonian (see e.g. Benjamin 1984, Shepherd 1990).

Ideally one would like to be able to satisfy both requirements,
Hamiltonian structure and formal accuracy. However, so far
they have proved to be incompatible beyond leading order. Arguments
can be mustered for one approach or the other. We do not wish to
enter that debate here, but simply choose to follow the
Hamiltonian approach, guided by scaling arguments, in what follows
--- recognizing its possible limitations.

The system we shall investigate is rotating shallow-water
flow. This is chosen because it is arguably the simplest
system in geophysical fluid dynamics that contains the essential
physics of nonlinear vortical motion coupled to fast (in this
case inertia-gravity) waves. Yet the approach may be expected to
have implications for other systems having the same essential
characteristics, such as vortex/acoustic-wave interactions in
compressible flow. Farge \& Sadourny (1989) proposed a weak-wave
model for shallow-water flow, but their derivation was rather {\it
ad hoc} and their system has a number of unsatisfactory features.
We are aware of no other study along these lines.

The plan of the paper is as follows. In Section 2 the properties
of the shallow-water equations are briefly reviewed. Our
weak-wave model is derived in Section 3, in a way that guarantees
its Hamiltonian character. We then describe some of its properties:
symmetries and conservation laws (Section 4); pseudoenergy and
pseudomomentum invariants and nonlinear stability (Section 5);
rigorous upper bounds on gravity-wave emission from unstable
parallel vortical shear flows (Section 6); linear eigenvalues and
instability (Section 7). Some of the limitations of our approach
are discussed in Section 8.

\bigskip
\noindent
{\bf 2. The shallow-water system}
\bigskip
The shallow-water equations, describing the dynamics of a
shallow, homogeneous, rotating fluid, are given by (e.g. Pedlosky
1987)
$$
\v_t + ( \v \cdot \bn ) \v + f \hat\z \times \v = - g \bn h
,\qquad h_t + \bn \cdot ( h \v ) = 0 ,\eqno(2.1)
$$
where $\v (x,y,t)$ is the (horizontal) velocity, $h(x,y,t)$ the
fluid depth, $g$ the (constant) gravitational acceleration, $f$ the
(constant) Coriolis parameter, $\hat\z$ the unit vertical vector, and $\bn
\equiv ( \pa / \pa x , \pa / \pa y ) \equiv ( \pa_x , \pa_y )$.
We assume either an unbounded domain with appropriate decay
conditions at infinity, or a doubly-periodic domain, or a
combination of the two (i.e. unbounded in one direction and
periodic in the other). This system has the following integral
invariants: energy $\H = \int {\textstyle{1\over 2}} \bigl( h | \v
|^2 + g h^2 \bigr) \, d\x $; zonal absolute momentum $\M = \int h (
u - fy) \, d\x$, where $\v \equiv (u,v)$; angular momentum $\J =
\int h [ xv -yu + {f \over 2} ( x^2 + y^2 ) ] \, d\x$; and a family
of Casimirs $\C = \int h C(q) \, d\x$ where $C( \cdot )$ is an
arbitrary function and $q = [f + \hat\z \cdot (\bn \times \v)]/h$
is the potential vorticity. The Casimir invariants are a
consequence of the material conservation of potential vorticity,
$q_t + \v \cdot \bn q = 0$, which follows from (2.1).

The system (2.1) can be written in the generic Hamiltonian form
(e.g. Shepherd 1990)
$$
\sigma_t = J {\delta \H \over \delta \sigma} ,\eqno(2.2)
$$
where $\sigma = ( \v , h )^T$ is the state vector of the
dynamical variables, $\H$ is the Hamiltonian, and the
cosymplectic form $J$ is given by
$$
J = \pmatrix{
0&q&-\pa_x\cr{}\cr-q&0&-\pa_y\cr{}\cr-\pa_x&-\pa_y&0\cr }
. \eqno(2.3)
$$

If one linearizes the
dynamics (2.1) about the rest state $\v = 0$, $h = H =
\hbox{const.}$, the dispersion relation is third order: one
root has frequency $\omega = 0$, corresponding to the vortical
motion; and there is a pair of roots with frequency $\omega =
\pm f \sqrt{1 + gH \kappa^2 / f^2 L^2}$, where $L$ is a horizontal
length scale and $\kappa$ the $O(1)$ dimensionless wavenumber,
corresponding to inertia-gravity waves. Allowing weak nonlinearity,
the vortical motion has a timescale $L/V$, where $V$ is a typical
velocity. The ratio of the inertia-gravity wave timescale to the
vortical timescale is thus $\epsilon \equiv V / |\omega | L = RB/
\sqrt{1 + B^2}$, where $R \equiv V/fL$ is the Rossby number and $B
\equiv  fL/\sqrt{gH}$. Evidently there is a separation in timescales
when $\epsilon \ll 1$. If one further assumes that the timescale of
the dynamics is advective (so that the gravity waves are filtered),
one obtains balanced dynamics: quasi-geostrophic dynamics for $R \ll
1$ with $B$ arbitrary, and barotropic dynamics for $B \ll 1$ with
$R = O(1)$ (or equivalently $F \ll 1$,
where $F \equiv V/\sqrt{gH}$ is the Froude number).

It may be noted that the energy of the full shallow-water equations
(2.1) is cubic in the dependent variables, which is a
consequence of the fact that the flow is divergent. (The same
situation arises with compressible flow.) This presents some serious
difficulties with respect to theoretical investigations. First,
the partitioning of energy between different scales of motion is
ambiguous, which is a problem for studies of
turbulence: Warn (1986), Farge \& Sadourny (1989), and Yuan \&
Hamilton (1994) have all had to invoke the
approximation of a quadratic energy in their analyses of
shallow-water turbulence, which is essentially a weak-wave
approximation. Second, stability theorems
based on the pseudomomentum and pseudoenergy, while easily derived
for small-amplitude disturbances (Ripa 1983), do not appear to
extend to finite amplitude (Shepherd 1992).

What we seek, therefore, is a weak-wave model for
shallow-water flow that includes the following features: quadratic
energy; Hamiltonian structure; appropriate conservation laws;
quasi-geostrophic or barotropic vortical motion; inertia-gravity
waves; and non-trivial coupling between the vortical motion and the
inertia-gravity waves.

\bigskip\noindent
{\bf 3. A Hamiltonian weak-wave model}
\bigskip
We begin by transforming the dependent variables in a way
that makes the vortical and wave motion more transparent in the
Hamiltonian structure. In particular, choose $\tilde q = \hat\z
\cdot ( \bn \times \v ) - f \eta / H$, $\Delta = \bn \cdot \v$,
and $\eta = h - H$ as the new dependent variables.
Writing $\v = \hat\z \times \bn \psi + \bn \chi$, we  now
non-dimensionalize the variables according to $
\psi = VL \psi_*$, $\chi = VL \chi_*$, $\tilde q = (V/L)
\tilde q_*$, $\bn = (1/L) \bn_*$, $t = (b/f) t_*$, and $\eta = HR
b \eta_*$, where $b \equiv B/\sqrt{1 + B^2}$. Note that $\tilde
q_* = \nabla^2 \psi_* - b \eta_*$, $\epsilon = Rb$, and $b \leq 1$.
Dropping the asterisks from now on, the resulting system can again
be cast in the form (2.2) with the state vector $\sigma = ( \tilde q
, \Delta , \eta )^T$, with $J$ given by
$$
J = \pmatrix{
- \epsilon \, \pa \bigl( { \tilde q \over 1 + \epsilon
\eta} , ( \cdot ) \bigr)& \epsilon \, \bn \cdot \bigl( { \tilde q
\over 1 + \epsilon \eta} \bn ( \cdot ) \bigr)&0\cr{}\cr - \epsilon
\, \bn \cdot \bigl( { \tilde q \over 1 + \epsilon \eta} \bn (
\cdot ) \bigr)& - \epsilon \, \pa \bigl( { \tilde q \over 1 +
\epsilon \eta} , ( \cdot ) \bigr)&- \nabla^2  ( \cdot )\cr{}\cr
0&\nabla^2
 ( \cdot )&0\cr} ,\eqno(3.1)
$$
where $\pa ( f,g) \equiv f_x g_y - f_y g_x$, and
with the Hamiltonian
$$
\H = \int {1\over 2} \Bigl\{ (1 + \epsilon
\eta ) ( | \bn \psi |^2 + | \bn \chi |^2 ) + 2 \epsilon \eta
\, \hat\z \cdot \bn \psi \times \bn \chi + {\eta^2 \over 1 + B^2}
\Bigr\} \, d\x .\eqno(3.2)
$$
(Note that in obtaining (3.2), the Casimir invariant $\int ( {1\over
2} g H^2 + gH \eta ) \, d\x$ has been removed from the original
Hamiltonian.) To this point no approximation has been made.

We now invoke the assumption of a timescale separation,
$\epsilon \ll 1$. To leading order, we
simply take $\epsilon = 0$ which yields
$$
J = \pmatrix{ 0&0&0\cr{}\cr0&0& -\nabla^2 ( \cdot )\cr
{}\cr 0&\nabla^2 ( \cdot )&0\cr} ,\qquad
\H = \int {1\over 2} \Bigl\{  | \bn \psi |^2 + |
\bn \chi |^2 ) + {\eta^2 \over 1 + B^2} \Bigr\} \, d\x
,\eqno\hbox{(3.3a,b)}
$$
with
$$
{\delta \H \over \delta \tilde q} = - \psi ,\qquad
{\delta \H \over \delta \Delta} = - \chi ,\qquad
{\delta \H \over \delta \eta} = {\eta \over 1 + B^2} - b \psi
.\eqno\hbox{(3.3c)}
$$
The system (3.3) describes the linearized dynamics
$$
{\pa \tilde q \over \pa t} = 0 ,\qquad
{\pa \Delta \over \pa t} = b \nabla^2 \psi - {\nabla^2 \eta \over
1 + B^2} , \qquad {\pa \eta \over \pa t} = - \Delta ,\eqno(3.4)
$$
with no vortical evolution. We now ask: what is the {\it simplest}
modification that one can make to (3.3) to include the
nonlinear vortical dynamics, while retaining the Hamiltonian
structure? Keeping all the $O(\epsilon )$ terms in (3.1) is
not an option, because Jacobi's identity would then not
hold: this is a typical problem in non-canonical Hamiltonian
perturbation expansions. However, we may now invoke our second
scaling assumption, namely that the divergence is weak
compared with the vorticity: $| \chi | \ll | \psi |$. Note that
the first column of (3.1) acts on $\delta \H / \delta \tilde q$,
which to leading order [see (3.3c)] is given by $-\psi$; while the
second column of (3.1) acts on $\delta \H / \delta \Delta$, which to
leading order [see (3.3c)] is given by $-\chi$. Thus we keep the
dominant term in each of the three equations by adding the
leading-order part of the upper-left entry of (3.1) to (3.3a),
namely
$$
J = \pmatrix{ - \epsilon \,\pa \bigl( \tilde q , (
\cdot ) \bigr)&0&0\cr{}\cr0&0& -\nabla^2 ( \cdot )\cr {}\cr
0&\nabla^2 ( \cdot )&0\cr} ,\eqno(3.5)
$$
while keeping the quadratic Hamiltonian (3.3b). It may be
verified that (3.5) satisfies all required properties of
cosymplectic forms, including Jacobi's identity. Note, however, that
although the gravity-wave motion must be weak, it cannot be too
weak or the divergence equation will be badly approximated by the
above procedure. In particular, in the divergence equation the
term arising from the first column of (3.1) must be negligible
against that arising from the third column; using the fact that
the dimensionless gravity-wave frequency and $\tilde q$ are both
$O(1)$, this requires $\epsilon | \psi | \ll | \chi |$. The formal
scaling regime for the validity of this system is therefore:
$$
\epsilon \ll 1,\qquad \epsilon | \psi | \ll | \chi | \ll |
\psi | .\eqno(3.6)
$$
If we absorb the factor $\epsilon$ in the nonlinear term by
rescaling all the variables by $1/\epsilon$, then (3.3b,c) and (3.5)
yields the system
$$
{\pa \tilde q \over \pa t} = - \pa ( \psi , \tilde q ) ,\qquad
{\pa \Delta \over \pa t} = \Bigl( {B^2 - \nabla^2 \over
1 + B^2} \Bigr) \eta + b \tilde q , \qquad {\pa \eta \over \pa t} =
- \Delta .\eqno\hbox{(3.7a,b,c)}
$$
This, together with the definition $\tilde q = \nabla^2 \psi
- b \eta$, is our weak-wave model.

The system (3.7) contains both nonlinear vortical dynamics
[through (3.7a)] and inertia-gravity wave dynamics [through
(3.7b,c)]. The coupling between the wave and vortical dynamics is
seen to come through the linear forcing term $b \tilde q$ in the
$\pa \Delta / \pa t$ equation, and through feedback from the
wave dynamics onto the $\pa \tilde q / \pa t$ equation
through $\psi$. The system is analogous to the reduced
nine-component low-order model of Lorenz (1986, Eqn.(2)), which also
has all the nonlinearity in the slow equation.

It is interesting to compare the system (3.7) with the
weak-wave model of Farge \& Sadourny (1989, Eqn.(15)); the latter
is given by
$$
{\pa \tilde q \over \pa t} = - \pa ( \psi , \tilde q ) ,\quad
{\pa \Delta \over \pa t} =  \Bigl( {B^2 - \nabla^2 \over
1 + B^2} \Bigr) \eta + b \tilde q - \pa ( \chi , \nabla^2 \psi ),
\quad {\pa \eta \over \pa t} = - \Delta - \pa ( \psi ,
\eta ).\eqno(3.8)
$$
Evidently (3.8) has additional nonlinear terms, but it is not
clear how they arise from the original shallow-water
system via (3.1). In addition, (3.8) does not appear to be
Hamiltonian.

\bigskip\noindent
{\bf 4. Symmetries and conservation laws}
\bigskip
In this section we discuss some of the dynamical implications of
the Hamiltonian structure of our weak-wave system. The reader
may wish to consult Shepherd (1990) for further background on
some of the points mentioned here.

The
approximation leading to (3.7) consists of keeping the leading
order (quadratic) approximation to the full Hamiltonian (3.2),
together with the simplest approximation to the full
cosymplectic form (3.1) that incorporates the vortical
dynamics yet still satisfies the requisite properties, including
Jacobi's identity. The resulting
weak-wave system (3.7) is thereby Hamiltonian by construction.
The cosymplectic form (3.5) represents the direct product of the
non-canonical quasi-geostrophic $J$, namely $ - \pa ( \tilde q , (
\cdot ))$, and the canonical linear-wave $J$. This kind of
structure can be expected to be generic for geophysical fluid
dynamical systems involving both fast and slow degrees of freedom.

Hamiltonian
structure provides a clear connection between symmetries and
conservation laws, via Noether's theorem,
and allows for the identification of Casimir invariants. The
Casimir invariants $\C$ are the solutions to $J ( \delta \C /
\delta \sigma ) = 0$; in the case of (3.5), this condition leads to
$$
{\delta \C \over \delta \tilde q} = \hbox{function of } \tilde
q, \qquad {\delta \C \over \delta \Delta} = \hbox{const.} ,\qquad
{\delta \C \over \delta \eta} = \hbox{const.},\eqno(4.1)
$$
which implies
$$
\C = \int \bigl\{ C ( \tilde q ) + c_1 \Delta + c_2 \eta
\bigr\} \, d\x .\eqno(4.2)
$$
Here $C ( \cdot )$ is an arbitrary function and $c_1$, $c_2$ are
arbitrary constants. With the exception of the second term,
which vanishes anyway, the Casimir invariants (4.2) are obvious
analogues of those present in the original shallow-water system,
except that there is no factor of $h$ in front of $C( \tilde q)$
(this reflects the fact that the advection of $\tilde q$ in
(3.7a) is solely by the non-divergent flow); also, $\int C(
\tilde q) \, d\x$ is identical to the Casimir invariants found in
quasi-geostrophic dynamics.

Noether's theorem (see e.g. Benjamin 1984) states that if $\delta
\sigma$ is an infinitesimal symmetry perturbation under which the
Hamiltonian is invariant, then a functional $\M$ satisfying
$\delta \sigma = J ( \delta \M / \delta \sigma )$ is a conserved
quantity. We consider both zonal translational symmetry and
rotational symmetry. In the first case, this condition leads
to
$$
- \pa \Bigl( \tilde q , {\delta \M \over \delta \tilde q} \Bigr)
= - \tilde q_x, \qquad - \nabla^2 {\delta \M \over \delta
\eta} = - \Delta_x ,\qquad \nabla^2 {\delta \M \over \delta
\Delta} = - \eta_x , \eqno(4.3)
$$
which may be shown to imply (within a Casimir, of course)
the zonal momentum invariant
$$
\M = \int \Bigl( y \tilde q + \eta {\pa \chi \over \pa x} \Bigr)
\, d\x .\eqno(4.4)
$$
In the case of rotational symmetry with respect to an angle
$\theta$, Noether's theorem leads to the angular momentum
invariant $$
\J = \int \Bigl( - {1\over 2} r^2 \tilde q + \eta {\pa \chi \over
\pa \theta} \Bigr) \, d\x .\eqno(4.5)
$$
The terms involving $\tilde q$ in (4.4) and (4.5)
represent an approximation to the momentum and angular momentum of
the rotational part of the flow, in vorticity form; they are
identical to the momentum and angular momentum invariants of
quasi-geostrophic dynamics. The terms involving $\chi$ in (4.4)
and (4.5) represent the momentum and angular momentum of the
divergent part of the flow.

\bigskip\noindent
{\bf 5. Pseudoenergy, pseudomomentum, and nonlinear stability}
\bigskip

A general property of Hamiltonian systems is that any
basic state possessing a symmetry invariance is a
constrained extremal of the integral invariant corresponding to
that symmetry, the constraint being imposed through a suitable
Casimir. This means that one may construct exact, finite-amplitude
invariants that are quadratic to leading order in the disturbance
to such basic states. For example, with steady basic states
one can combine the energy and Casimir invariants to obtain a
so-called pseudoenergy invariant. In the same way, with
zonally symmetric basic states one can combine the zonal
momentum and Casimir invariants to obtain a pseudomomentum
invariant. Such invariants can be used to derive nonlinear
stability theorems. See Shepherd (1990, 1994) for further
background.

\bigskip\noindent
{\it (a) Zonally symmetric basic states}
\bigskip
We first consider disturbances $\{ \tilde q' , \Delta' , \eta' \}$
to a steady, zonally symmetric basic state $\{ \tilde q_0 (y) ,
\Delta_0 (y) , \eta_0 (y) \}$. In this case we can use the
energy (3.3b), Casimir (4.2), and zonal momentum (4.4)
invariants to construct the combined invariant
$$
\A \equiv \bigl( \H + \C + \alpha \M \bigr) [ \tilde q_0 +
\tilde q' , \Delta_0 + \Delta' , \eta_0 + \eta' ] -
\bigl( \H + \C + \alpha \M \bigr) [ \tilde q_0 , \Delta_0 ,
\eta_0 ] ,\eqno(5.1)
$$
where $\alpha$ is kept free. The Casimir $\C$ is determined by
the extremal condition $\delta \A = 0$, which leads to
$$
C' ( \tilde q_0 ) = \psi_0 - \alpha y ,\qquad
\Delta_0 = 0 ,\qquad \eta_0 = b (1 + B^2 ) \psi_0
.\eqno\hbox{(5.2a,b,c)}
$$
(Note that $c_1$ and $c_2$ can be set to zero without loss of
generality by imposing a suitable gauge condition on the
potentials $\psi_0$ and $\chi_0$.) Conditions (5.2b,c) express
geostrophic balance, which is required for a steady solution.
Condition (5.2a) determines the functional dependence of $C (
\cdot )$; thus the right-hand side of (5.2a) must be interpreted
as a function of $\tilde q_0$. If we introduce the function
$G_{\alpha} ( \cdot )$ defined by $G_{\alpha} ( \tilde q_0
(y) ) = \psi_0 (y) - \alpha y$, then $\A$
can be written
$$
\eqalignno{
\A &= \int \biggl\{ {1\over 2} \biggl( | \bn \psi' |^2 + | \bn
\chi' |^2 + { ( \eta ')^2 \over 1 + B^2} \biggr) - \psi_0 \tilde
q' + \int_{\tilde q_0}^{\tilde q_0 + \tilde q'} \!\! G_{\alpha} (
\xi ) \, d \xi + \alpha y \tilde q' + \alpha \eta' {\pa \chi' \over
\pa x} \biggr\} \, d\x \cr &= \int \biggl\{ {1\over 2} \biggl( |
\bn \psi' |^2 + \Bigl( {\pa \chi' \over \pa y} \Bigr)^2 + \Bigl( {
\eta ' \over \sqrt{1 + B^2}} + \alpha \sqrt{1 + B^2} {\pa \chi'
\over \pa x} \Bigr)^2 + \bigl[ 1 - \alpha^2 (1 + B^2 ) \bigr]
\Bigl( {\pa \chi' \over \pa x} \Bigr)^2 \biggr) \cr
&\qquad +
\int_{0}^{\tilde q'} \Bigl[ G_{\alpha} ( \tilde q_0 + \xi ) -
G_{\alpha} ( \tilde q_0 ) \Bigr] d \xi \biggr\} \, d\x .&(5.3)\cr }
$$
This expression is seen to be positive definite if
$$
( U_0 + \alpha ) {d \tilde q_0 \over dy} < 0 \quad \forall y
\qquad \hbox{and} \qquad \alpha^2 < {1 \over 1 + B^2}
,\eqno\hbox{(5.4a,b)}
$$
using $dG_{\alpha}/d\tilde q_0 = [ dG_{\alpha}/dy ] / ( d\tilde q_0 / dy ) = -
( U_0 + \alpha ) / ( d\tilde
q_0 / dy )$, where $U_0 \equiv - d\psi_0 / dy$. (Note that
under the condition (5.4a), $G_{\alpha} ( \cdot )$ is
guaranteed to be single-valued.) Since $\A$ is an exact invariant of
the nonlinear dynamics, it follows using standard methods that
whenever $\alpha$ can be chosen such that (5.4a,b) are satisfied,
the basic state is nonlinearly stable. In particular, a normed
stability theorem can be established\footnote{\dag}{Technically
this requires that the left-hand side of (5.4a) be bounded away
from zero and minus infinity.}, which holds for disturbances of
{\it arbitrarily} large magnitude.

It was Arnol'd (1966) who first introduced this approach
[building on earlier work of Fj{\o}rtoft (1950)] and applied it
to the barotropic system. For that system, (5.4a) alone is a
sufficient condition for nonlinear stability. The same result
applies to the quasi-geostrophic system. Our stability theorem is
thus seen to be the analogue of Arnol'd's theorem for the
weak-wave model, with the additional condition (5.4b) arising
from the model's gravity wave (unbalanced) dynamics.

Ripa (1983) followed the above methodology for the full
shallow-water system, and showed that the second variation of
the corresponding invariant $\A$ was positive definite if an
$\alpha$ could be found such that (in terms of our present
variables)
$$
( U_0 + \alpha ) {d q_0 \over dy} < 0 \quad \forall y
\qquad \hbox{and} \qquad ( \alpha + U_0 )^2 < {1 + \eta_0 \over 1
+ B^2} \quad \forall y ,\eqno\hbox{(5.5a,b)}
$$
where $q_0 = (b + \nabla^2 \psi_0 ) / ( 1 + \eta_0 )$ is a
scaled potential vorticity. Note that under the rescaling implicit
in (3.7), $| \eta_0 | = O(\epsilon ) \ll 1$ so $q_0 \approx b +
\tilde q_0$ and the right-hand side of (5.5b) reduces to the
right-hand side of (5.4b). Thus the stability
theorem resulting from conservation of $\A$ in the weak-wave model
is analogous to that in the full shallow-water model. The key
difference is that the higher-order terms in the shallow-water form
of $\A$ cannot be controlled (it appears that only the second
variation of $\A$ can be shown to be positive definite), so
condition (5.5) is only a statement of small-amplitude (linear)
stability.

The fact that the weak-wave stability theorem is fully nonlinear
means that it can be used to obtain rigorous upper bounds on the
nonlinear saturation of instabilities in this model, following the
method of Shepherd (1988). In Section 6 it is shown how this
technique places rigorous bounds (within the context of this
weak-wave model) on the emission of gravity waves from
unstable parallel vortical flows.

\bigskip\noindent
{\it (b) Axisymmetric basic states}
\bigskip
We now consider disturbances
to a steady axisymmetric basic state $\{ \tilde q_0 (r) ,
\Delta_0 (r) , \eta_0 (r) \}$ where $x = r \cos\theta$, $y =
r\sin\theta$. In this case we proceed exactly as above, but
using the angular momentum invariant (4.5) in place of the zonal
momentum. Defining $\P$ according to (5.1) but with $\M$
replaced by $\J$, the extremal condition $\delta \P = 0$
leads to
$$
C' ( \tilde q_0 ) = \psi_0 + \alpha {r^2 \over 2} ,\qquad
\Delta_0 = 0 ,\qquad \eta_0 = b (1 + B^2 ) \psi_0
.\eqno\hbox{(5.6a,b,c)}
$$
Introducing the function
$G_{\alpha} ( \cdot )$ defined by $G_{\alpha} ( \tilde
q_0 (r) ) = \psi_0 (r) + \alpha r^2/2$, it follows that $\P$ can be
written
$$
\eqalignno{
\P &= \int \biggl\{ {1\over 2} \biggl( | \bn \psi' |^2 + | \bn
\chi' |^2 + { ( \eta ')^2 \over 1 + B^2} \biggr) - \psi_0 \tilde
q' + \int_{\tilde q_0}^{\tilde q_0 + \tilde q'} \!\! G_{\alpha} ( \xi ) \, d
\xi - \alpha {r^2 \over 2}
\tilde q' + \alpha \eta' {\pa \chi' \over \pa \theta} \biggr\}
\, d\x \cr
&= \int \biggl\{ {1\over 2} \biggl( | \bn \psi' |^2 + \Bigl(
{\pa \chi' \over \pa r} \Bigr)^2 + \Bigl( { \eta ' \over \sqrt{1
+ B^2}} + \alpha \sqrt{1 + B^2} {\pa \chi' \over \pa \theta}
\Bigr)^2 + \Bigl[ {1 \over r^2} - \alpha^2 (1 + B^2 ) \Bigr]
\Bigl( {\pa \chi' \over \pa \theta} \Bigr)^2 \biggr) \cr
&\qquad +
\int_{0}^{\tilde q'} \Bigl[ G_{\alpha} ( \tilde q_0 + \xi ) -
G_{\alpha} ( \tilde q_0 ) \Bigr] d \xi \biggr\}
\, d\x .&(5.7)\cr }
$$
As with $\A$, $\P$ is an exact invariant of
the nonlinear dynamics. The expression (5.7) is seen to be positive
definite if $$
( U_0 + \alpha r) {d \tilde q_0 \over dr} > 0 \quad \forall r
\qquad \hbox{and} \qquad \alpha^2 < {1 \over r^2 (1 + B^2 )}
,\eqno\hbox{(5.8a,b)}
$$
where $U_0 \equiv d \psi_0 / dr$. However, it turns out that for
physically realizable flows in an unbounded domain, (5.8a,b) can
{\it never} be satisfied! To see this, first note that taking $r
\to\infty$ in (5.8b) implies that $\alpha = 0$. Using (5.6c),
$\tilde q_0 = d^2 \psi_0 / dr^2 + (1/r) d\psi_0 / dr - B^2
\psi_0$. Therefore (5.8a) requires
$$
\eqalignno{
0 < \int_0^{\infty} &{d \psi_0 \over dr} \, {d \tilde q_0 \over
dr} \, dr
=\int_0^{\infty} \Bigl\{ {d \psi_0 \over dr} \, {d^3 \psi_0 \over
dr^3} + {d \psi_0 \over dr} \, {d\over dr} \Bigl( {1\over r} \,
{d \psi_0 \over dr} \Bigr) - B^2 \Bigl( {d \psi_0 \over dr}
\Bigr)^2 \Bigr\} \, dr \cr
&= - \int_0^{\infty} \Bigl\{ \Bigl( {d^2 \psi_0
\over dr^2} \Bigr)^2 + B^2 \Bigl( {d \psi_0
\over dr} \Bigr)^2 \Bigr\} \, dr + \int_0^{\infty}
{d \psi_0 \over dr} \, {d\over dr} \Bigl( {1\over r} \,
{d \psi_0 \over dr} \Bigr) \, dr ,&(5.9) \cr}
$$
using the conditions $dU_0 /dr \to 0$ as $r
\to \infty$ and $U_0 \to 0$ as $r \to 0$. But
$$
\eqalignno{
\int_0^{\infty}
{d \psi_0 \over dr} \, {d\over dr} \Bigl( {1\over r} \,
&{d \psi_0 \over dr} \Bigr) \, dr = \int_0^{\infty} \Bigl\{
{1\over r} \, {d \psi_0 \over dr} \, {d^2 \psi_0 \over dr^2}
- {1\over r^2} \, \Bigl( {d \psi_0 \over dr} \Bigr)^2 \Bigr\} \,
dr \cr
&= - \int_0^{\infty}
{d \psi_0 \over dr} \, {d\over dr} \Bigl( {1\over r} \,
{d \psi_0 \over dr} \Bigr) \, dr - \int_0^{\infty}
{1\over r^2} \, \Bigl( {d \psi_0 \over dr} \Bigr)^2 \, dr
,&(5.10)\cr}
$$
since $U_0 = O(r)$ as $r \to 0$ for finite vorticity. Hence
the second integral on the right-hand side of (5.9) is negative
definite, which implies that (5.8) can never be satisfied.

In the barotropic or quasi-geostrophic case, (5.8a) alone is a
sufficient condition for nonlinear stability, with no
restriction on $\alpha$. In the shallow-water case, even a
small-amplitude stability theorem cannot be obtained by this
approach for unbounded domains (see Appendix). So we conclude that
the weak-wave model captures the essential physics of the full
shallow-water dynamics with regard to stability theorems of the
Arnol'd type.

\bigskip\noindent
{\bf 6. Rigorous upper bounds on gravity-wave emission from
unstable parallel shear flows}
\bigskip
Suppose we are given a parallel shear flow $\{ \tilde q_1 (y) ,
\eta_1 (y) \}$ which is known to be unstable. In general we expect
the instability, even if it is vortical in nature, to excite
gravity waves. For example, Ford (1994$b$) has studied shear-layer
(Rayleigh-type) instability
in the shallow-water equations, and has found the generation of
gravity waves to be ubiquitous. An important question is
whether there are any {\it a priori\/} upper bounds on the
amplitude of such gravity-wave emission.

Such bounds can be obtained using our nonlinear stability theorem.
Following Shepherd (1988), we may regard the
initial unstable flow as a finite-amplitude disturbance to some
nearby stable flow, and then invoke the nonlinear stability
theorem. Since steady flows are necessarily balanced (with no
gravity-wave component), any bound on the disturbance will also
provide a bound on the gravity-wave activity. We provide an
explicit demonstration of this by deriving a bound on the
kinetic energy of the divergent flow,
$$
\E_d \equiv \int {1\over 2} | \bn
\chi |^2  \, d\x .\eqno(6.1)
$$
Let us assume, for the sake of simplicity, that the unstable flow
is perturbed infinitesimally at $t=0$. (It would be
straightforward to include the effect of a finite initial
perturbation.) Now, introduce a basic state $\{ \tilde q_0 (y) ,
\eta_0 (y) \}$ that is stable by criteria (5.4). At $t=0$, the
disturbance $\{ \tilde q' , \Delta' , \eta' \}$ relative to this
basic state is given by $\eta'(0) \approx \eta_1 - \eta_0$, etc.,
and the invariant $\A$ is approximately given by
$$
\A (0) = \int \biggl\{ {1\over 2} \biggl[ \Bigl( {d (\psi_1
- \psi_0 ) \over dy} \Bigr)^2 + { ( \eta_1 - \eta_0 )^2 \over 1 +
B^2} \biggr] +
\int_{\tilde q_0}^{\tilde q_1} \Bigl[ G_{\alpha} ( \xi ) -
G_{\alpha} ( \tilde q_0 ) \Bigr] d \xi \biggr\} \, d\x .\eqno(6.2)
$$
For $t>0$, we note that $\chi = \chi'$ (since the basic state has
no divergent component); then from (6.1) and (5.3), together with
the conservation of $\A$ in time, we obtain the inequality
$$
\E_d (t) \leq {\A (t) \over 1 - \alpha^2 (1 + B^2 )} =
{\A (0) \over 1 - \alpha^2 (1 + B^2 )} ,\eqno(6.3)
$$
into which (6.2) may be substituted. (Note that the factor $1 -
\alpha^2 (1 + B^2 )$ is always positive, by virtue of (5.4b).)
In this way we obtain an {\it
a priori\/} upper bound on the divergent kinetic energy $\E_d (t)$
for all time, which is a functional of the initial unstable flow
and of the chosen stable basic state (including the choice of
$\alpha$). This places a rigorous bound on the
emission of gravity waves from the unstable flow. Since the basic
state is arbitrary, one may try to minimize the bound over all
possible choices of stable basic state (cf. Shepherd 1988). In the
context of the present weak-wave model, this analysis has been
performed by Nore (1994) for the Bickley jet and the tanh$(y)$
shear layer.

For an initial flow that is marginally unstable, the above
procedure yields a bound that is a function of the
supercriticality --- implying weak gravity-wave emission. Arguments
from statistical mechanics (Warn 1986) would suggest, to the
contrary, that once a steady flow goes unstable, its energy will
be partitioned roughly equally into both vortical and
gravity-wave components. The nonlinear stability theorem therefore
provides definite limits on the extent of such
``thermalization'' in the weak-wave model (cf. Shepherd 1987).

\bigskip\noindent
{\bf 7. Linear eigenvalues and instability}
\bigskip
The shallow-water equations allow mixed vortical/gravity-wave
instabilities that are filtered by the quasi-geostrophic
equations. An important test of our weak-wave model is the
extent to which it can capture such instabilities. As a
preliminary step, we consider the instability properties of two
simple basic flows.

\bigskip\noindent
{\it (a) Zonally symmetric basic state}
\bigskip
We first consider the case of a piecewise-constant
zonally symmetric potential vorticity distribution with a single
jump at $y=0$. As such a flow
has no intrinsic length scale, we may take $B=1$ without loss of
generality, and introduce the basic state
$$
\tilde q_0 = \cases{2Q,&$y>0$\cr {}\cr 0,&$y<0$\cr} \quad
,\qquad \psi_0 = \cases{-2Q+Q \e^{-y},&$y>0$\cr {}\cr
\qquad -Q\e^{y},&$y<0$\cr} \quad ,\eqno(7.1)
$$
with $\eta_0 = \sqrt{2} \psi_0$. Again without loss of
generality, we take $Q > 0$. The velocity field associated with
this basic state is given by $U_0 (y) = Q \e^{-|y|}$, namely an
eastward symmetric jet with a maximum at $y=0$.

Physically this problem consists of gravity waves in the upper
and lower half-planes, coupled to a potential-vorticity
disturbance at $y=0$. In each of the two half-planes we
introduce a normal-mode form for $\psi'$, proportional to
$\e^{i(kx + \ell y - \omega t)}$, and insist on evanescence as
$|y| \to \infty$. The frequency $\omega$ must of course be the same
in the two half-planes; the same condition applies to the (real)
zonal wavenumber $k$, by virtue of the kinematic condition that $v'
= \psi'_x$ must be continuous across $y=0$. Away from $y=0$, we have
$\tilde q' = 0$ and so (3.7b,c) --- with $B=1$ --- yield
$$
\omega^2 = {1\over 2} \bigl( 1 + k^2 + \ell^2 \bigr) ,\eqno(7.2)
$$
which is just the dispersion relation for inertia-gravity waves.
We obtain a dynamical matching condition at $y=0$ by
integrating the linearized potential vorticity equation (3.7a),
namely ${\tilde q'}_t + U_0 {\tilde q'}_x + \psi'_x \tilde
q_{0y} = 0$, across $y=0$. This yields $( \omega - U_0 (0) k) [
\ell ] = - ik [ \tilde q_0 ]$, where $[ \cdot ]$ denotes the
jump across $y=0$. Evidently a jump in $\ell$ is required to
balance the jump in $\tilde q_0$. Since (7.2) implies that
$\ell^2$ must be the same on either side of $y=0$, this means
that $\ell$ must change sign. Letting $\ell$ denote the value
for $y > 0$, the dynamical matching condition may be written
$$
( \omega - Qk ) \ell = - i Qk .\eqno(7.3)
$$
Equations (7.2) and (7.3) define the linear disturbance
problem. Note that $\ell_i > 0$ for evanescent disturbances; it
may be shown that this requires the intrinsic frequency
$\hat\omega \equiv
\omega - Qk$ to have a negative real part. For
quasi-geostrophic dynamics, the condition (7.2) is replaced by the
condition $\ell^2 = - k^2 - 1$, which when substituted into (7.3)
leads to the single (stable) solution branch
$$
\omega_{{\smrm QG}} = Qk - {Qk \over \sqrt{1 + k^2}}
.\eqno(7.4)
$$

Condition (5.4a) is satisfied (in the appropriate limiting sense
at $y=0$) if $\alpha < -Q$, so the basic state (7.1) is
provably stable by our criterion (5.4) for $Q < Q_c \equiv 1
/\sqrt{2}$. This stability condition has a simple physical
interpretation. From (7.2), the phase speed of the gravity-wave
mode is always greater than $1/\sqrt{2}$, but approaches
that value in the limit $k \to \infty$. From (7.4), the phase
speed of the quasi-geostrophic mode is always less than $Q$, but
approaches that value in the limit $k \to \infty$. So when $Q <
Q_c$, the phase speeds of the two modes are always
separate; but when $Q > Q_c$, they cross for
sufficiently large $k$. Our stability condition (5.4) is
therefore a condition for the absence of a mode crossing.

In terms of dimensional variables, the critical value corresponds
to $[ \tilde q ] = 2f$. This lies outside the formal domain of
validity of the weak-wave model, since it requires order-unity
relative changes in the layer depth.

Combining (7.2) and (7.3) leads to the
quartic equation
$$
2 \hat\omega^4 + 4Qk \hat\omega^3
+ ( [ 2 Q^2 - 1 ] k^2 - 1) \hat\omega^2 + Q^2 k^2 = 0.\eqno(7.5)
$$
We can analyze (7.5) asymptotically in the limits of small and
large wavenumber $k$. Note that the coefficient of $\hat\omega^2$
changes character depending on the value of $Q$: for $Q < Q_c$, it
is always negative; for $Q > Q_c$, it changes sign at
some value of $k$. This latter case corresponds, as anticipated, to
a mode crossing, as shown below.

For $k \ll 1$, the method of dominant balance
leads to the two stable solutions
$$
2 \hat\omega^4 \sim \hat\omega^2 \qquad
\Longrightarrow \qquad \hat\omega \sim - {1 \over \sqrt{2}} ,
\quad \omega \sim - {1 \over \sqrt{2}} \eqno(7.6)
$$
and
$$
 \hat\omega^2 \sim Q^2 k^2 \qquad
\Longrightarrow \qquad \hat\omega \sim - Qk ,\quad
\omega \sim {Qk^3 \over 2} .\eqno(7.7)
$$
In each case there are two real roots but we have kept only the
one with $\hat\omega < 0$. The root (7.6) evidently corresponds to
the small-$k$ limit of a pure gravity-wave mode (7.2), while
the root (7.7) corresponds to the small-$k$ limit of the
quasi-geostrophic mode (7.4). (The expression for $\omega$ in
(7.7) obtains on expanding (7.5) in $k$.)

For $k \gg 1$, the method of dominant balance leads to the
solutions
$$
2 \hat\omega^4 + 4Qk \hat\omega^3
\sim ( 1 - 2 Q^2 ) k^2
\hat\omega^2
\qquad \Longrightarrow \qquad \hat\omega \sim - Qk \pm {k
\over \sqrt{2}} ,\quad \omega \sim \pm {k
\over \sqrt{2}} \eqno(7.8)
$$
and
$$
( 1 - 2Q^2 ) k^2
\hat\omega^2 \sim Q^2 k^2 \qquad
\Longrightarrow \qquad \hat\omega \sim \pm {Q \over \sqrt{1 -
2Q^2}} , \quad \omega \sim Qk
.\eqno(7.9)
$$
There are two distinct cases here. When $Q < Q_c$,
then only the negative root of (7.8) has $\hat\omega < 0$; it is
evidently the large-$k$ limit of a pure gravity-wave mode (7.2).
In this case another stable solution is provided by the
negative root of (7.9); it reduces to the large-$k$ limit of the
quasi-geostrophic mode (7.4) for $Q \ll 1$, and generally
has the same characteristic of the intrinsic frequency
asymptoting to a constant. When $Q > Q_c$, on the
other hand, (7.9) implies an unstable solution with $\hat\omega_i
\neq 0$; in this case both roots of (7.8) have $\hat\omega < 0$.

Therefore, for $Q < Q_c$, when our
stability theorem predicts stability for this basic flow, we
find that both quasi-geostrophic and
gravity-wave neutral modes can be identified, and are connected
smoothly (i.e. arise from the same dominant balance) in the
small-$k$ and large-$k$ limits: there is no mode crossing. But for
$Q > Q_c$, there is a dramatic change in the
character of one of the solutions: the small-$k$ quasi-geostrophic
mode metamorphoses into a gravity-wave mode in the large-$k$
limit, and an unstable solution appears.

Ford (1993) has
studied the analogous basic state (with piecewise-constant
$q$ rather than piecewise-constant $\tilde q$) in the context of the
shallow-water equations, obtaining disturbance equations
analogous to (7.2)--(7.3) but with the full basic-state height
field entering the problem and with a Doppler shifting of the
gravity-wave frequency. (Both of these effects are neglected in
the weak-wave model.) Ford found a similar transition in the character of the
neutral
modes at a critical value of $Q$, corresponding to the critical
value predicted by Ripa's stability criterion (5.5); for $Q$
larger than the critical value, he found mode crossing and
instability. In terms of our non-dimensionalization, Ford's
critical value of $Q$ (based on the jump in relative vorticity) is
$3/(4\sqrt{2})$, which is somewhat less than our $Q_c$. Moreover,
in the shallow-water case the growth rates are exponentially
small for large $k$, whereas here they are of order unity. Both
these discrepancies reflect the fact that the instability occurs
outside the range of validity of the weak-wave model.

\vfill\eject
\noindent
{\it (b) Axisymmetric basic state}
\bigskip
We now consider the case of a piecewise-constant
axisymmetric potential vorticity distribution with a single jump
at $r=1$:
$$
\tilde q_0 = \cases{Q,&$r<1$\cr {}\cr 0,&$r>1$\cr} \quad
,\qquad B^2 \psi_0 = \cases{-Q+BQ K_1(B) I_0(Br),&$r<1$\cr {}\cr
\qquad -BQ I_1(B) K_0 (Br),&$r>1$\cr} \quad ,\eqno(7.10)
$$
with $\eta_0 = (1+B^2) b \psi_0$. Here $I_m$ and $K_m$ are
modified Bessel functions of the first and second kind, of order
$m$. The velocity field associated with this vortex is continuous in
$r$, and has a maximum at $r=1$ given by $U_0 (1) = Q K_1(B) I_1
(B)$.

As in the zonally symmetric case, this problem consists of
gravity waves inside and outside the vortex, coupled to a
potential vorticity disturbance at $r=1$. Away from $r=1$, we
introduce a normal-mode form for $\psi'$, proportional to
$f(r) \e^{i(m\theta - \omega t)}$ where $f(r) = I_m ( \gamma
r)$ for $r<1$ and $f(r) = I_m ( \gamma ) K_m ( \gamma r ) / K_m
( \gamma )$ for $r>1$ (so $f(r)$ is continuous across $r=1$).
Taking $\tilde q' = 0$ for $r \neq 1$, (3.7b,c) yield
the inertia-gravity wave dispersion relation
$$
(1 + B^2 ) \omega^2 = B^2 - \gamma^2 .\eqno(7.11)
$$
As above, we obtain a dynamical matching condition at $r=1$ by
integrating the linearized potential vorticity equation
across $r=1$. After some algebra, this can be shown to reduce to
$$
\omega = Qm \bigl[ K_1(B) I_1(B) - K_m ( \gamma ) I_m (\gamma )
\bigr] .\eqno(7.12)
$$
Equations (7.11) and
(7.12) define the linear disturbance problem. The
quasi-geostrophic problem consists of (7.12) alone with $\gamma$
replaced by $B$, which gives a single (stable) mode. In the
limit $B \to 0$, this solution reduces to Kelvin's (1880)
result $\omega = Q (m-1)/2$.

The absence of a stability theorem in the axisymmetric case
suggests the existence of mode crossing between the gravity-wave
and quasi-geostrophic modes for all parameter values. That this
is indeed the case can be seen as follows. In the limit $m \gg
\max \{B,1 \}$, (7.12) suggests that $\omega = O(m)$; then (7.11)
implies $\gamma \sim \pm i \sqrt{1 + B^2} \omega$. Since $K_m (
\gamma ) I_m ( \gamma ) = O ( 1/\gamma )$ for $\gamma \gg 1$,
the second term of (7.12) is much smaller than the first, which
is consistent with the assumption that $\omega = O(m)$.
Substituting this expression for $\gamma$ back into (7.12), an
imaginary component to $\omega$ is implied: in particular,
$$
\omega_r \sim Qm K_1 (B) I_1(B) ,\qquad \omega_i \sim \pm {\pi
\over 2} Qm \Bigl[ J_m \Bigl( \sqrt{Q^2 m^2 K_1^2(B) I_1^2(B)
(1+B^2 ) - B^2} \, \Bigr) \Bigr]^2 ,\eqno(7.13)
$$
where $J_m$ is the Bessel function of the first kind, of order
$m$. We thus find instability for sufficiently large $m$, for all
parameter values. In the limit of small $Q$ and large $m$, we may
make the substitution $\sech \beta = Q K_1 (B) I_1 (B)
\sqrt{1 + B^2}$ in (7.13), and use the asymptotic approximation
$$
J_m ( m \, \sech \beta ) \sim {\e^{m( \tanh\beta - \beta
)} \over \sqrt{2 \pi m \, \tanh\beta}} .\eqno(7.14)
$$
Together with the identity $\e^{2\beta} = (1 + \tanh\beta)/(1 -
\tanh\beta )$, this yields an approximate growth rate
$$
\omega_i \sim {Q \over 4} \, {\e^{2m\tanh\beta} \over \tanh\beta}
\, \Bigl( {1 - \tanh\beta \over 1 + \tanh\beta} \Bigr)^m
.\eqno(7.15)
$$
Now using the small-$Q$ expansion
$$
\tanh\beta = \sqrt{1 - \sech^2 \beta} \sim 1 - {1\over 2} \, Q^2
K_1^2 (B) I_1^2 (B) (1 + B^2 ) + \dots ,\eqno(7.16)
$$
we obtain
$$
\omega_i \sim {Q \over 4} \, \e^{2m} \, \Bigl(
{1\over 2} \, Q K_1 (B) I_1 (B) (1 + B^2 )^{1/2}
\Bigr)^{2m} .\eqno(7.17)
$$
The derivation of this expression, valid for small $Q$ and large
$m$, was provided by R. Ford (personal communication, 1996). The
Froude number based on the velocity at the vortex edge
is $F = (1 + B^2 )^{1/2} U_0(1)$, namely $F = Q K_1 (B)
I_1 (B) (1 + B^2 )^{1/2}$, so we see from (7.17) that the
growth rate scales as $F^{2m}$. This is in accord with the
instability analysis of Ford (1994$a$) for the analogous basic state
in the shallow-water equations (with piecewise-constant $q$
rather than piecewise-constant $\tilde q$); Ford obtained
$$
\omega_i \sim {1 \over 4} \, \e^{2(m-1)} \, \Bigl( {F \over 4}
\Bigr)^{2m} \eqno(7.18)
$$
in the limit of large $m$ and small $F$. The factor of $Q$
difference between (7.17) and (7.18) is accounted for by the
different nondimensionalization of timescales, and the factor of
$2^{2m}$ difference by the fact that Ford's vortex-edge velocity is
$1/2$ rather than $1$ (so that there is a factor of two difference
in the definition of $F$). The only remaining difference between
(7.17) and (7.18) is a factor of $\e^2$, the origin of which is
unclear.

When $B \ll 1$, then $K_1 (B) I_1 (B) (1 + B^2 )^{1/2} \to
1/2$ and the approximation (7.15) is still valid for $Q < 2$ (in
the large-$m$ limit). In this case, the weak-wave model predicts a
growth rate of $O(1)$ when $Q = O(1)$. This is very much larger
than the shallow-water growth rate of $O(F^{2m})$ (see (7.18)) for
$B \ll 1$ and $Q = O(1)$, and indicates that in this regime the
weak-wave model is outside its range of validity. The problem is
that although $B \ll 1$ for the vortex, the instability involves a
gravity wave with a very long wavelength, whose frequency is not
much larger than the rotation rate of the vortex. This illustrates
the pitfalls of a formal scaling analysis based on a single length
scale.

\bigskip\noindent
{\bf 8. Conclusion}
\bigskip
Atmospheric and oceanic flows can often be characterized as
consisting of slow nonlinear vortical motion evolving in the
presence of fast but weak linear-wave motion. We seek a reduced
dynamical model that can describe this situation, and that has the
following features: quadratic energy; Hamiltonian structure;
appropriate conservation laws; quasi-geostrophic or barotropic
vortical motion; fast waves; and non-trivial coupling between the
vortical motion and the fast waves. In this paper we have derived
such a ``weak-wave'' model in the context of the $f$-plane
shallow-water equations, where the fast waves are inertia-gravity
waves.

Although our model is appealing from the Hamiltonian perspective,
it has at least two shortcomings. The first is that our method of
approximation is not unique: if we had chosen different prognostic
(dynamical) variables, we would have obtained a somewhat different
system. This reflects the fact that our model is not based on a
rigorous perturbation expansion. Against this we can only say that
the problem of combining rigorous perturbation expansions with
non-canonical Hamiltonian structure remains an open problem, which
nobody has yet been able to solve.

The second shortcoming of our model is that, for formal
self-consistency, the fast waves must not be {\it too} weak. This
might cast doubt on the utility of the model for studying mixed
vortical/gravity-wave instabilities, and the breakdown of balanced
dynamics. To determine whether the model is deficient in these
respects, we have analyzed its stability and instability
properties.

With regard to stability, the model possesses an
Arnol'd-type stability theorem for zonally symmetric vortical
basic flows, (5.4), that corresponds to the analogous theorem (5.5)
for shallow-water dynamics, and that has the same physical
interpretation: a basic flow is stable if its velocity is
everywhere subcritical with respect to the gravity-wave phase speed.
However, in our case the stability theorem is fully nonlinear.
This fact permits the establishment of rigorous upper bounds on
gravity-wave emission from unstable parallel shear flows. As is
the case with shallow-water dynamics (see Appendix), an Arnol'd-type
stability theorem cannot be obtained for axisymmetric vortical
basic flows. This is in striking contrast with the
quasi-geostrophic case --- where any vortex with a monotonic
potential-vorticity profile is nonlinearly stable --- and reflects
the destabilizing role of gravity waves in our model.

With regard to instability, we have examined the linear dynamics
of disturbances to two piecewise-constant potential-vorticity
distributions. In the case of the zonally symmetric basic state,
corresponding to an exponential jet, the nature of the eigenvalue
problem depends on the jet strength. When the jet is subcritical,
the modes are everywhere neutral and can be clearly
identified with quasi-geostrophic and gravity-wave modes. When the
jet is supercritical, however, the eigenvalues collide (mode
crossing) and instability occurs for sufficiently short zonal
wavelengths. This behaviour corresponds qualitatively to
that seen in the shallow-water system (Ford, 1993). In the case of
the axisymmetric vortex, instability is found for all parameter
values: gravity waves destabilize a vortex that would be stable
under balanced dynamics. Once again, this is
qualitatively consistent with the behaviour of the shallow-water
system (Ford, 1994$a$). It is also consistent with
the study of Broadbent \& Moore (1979), who found a destabilization
of axisymmetric vortices by coupling to acoustic waves in the
context of the two-dimensional compressible equations.

The gravity-wave equation implied by (3.7) is
$$
{\pa^2 \Delta \over \pa t^2} + \Bigl( { B^2 - \nabla^2 \over 1 +
B^2} \Bigr) \Delta = - b \pa ( \psi , \tilde q ) = - b \pa ( \psi
, \nabla^2 \psi ) + b^2 \pa ( \psi , \eta ).\eqno(8.1)
$$
Although
the zeroth moment of the right-hand side of (8.1) clearly
vanishes (for a compact vorticity distribution), the first moment
does not appear to vanish. This is in contrast to the
shallow-water system, which has vanishing zeroth and first
moments (Ford {\it et al.} 1996),
and suggests that our weak-wave model might
overestimate the forcing of gravity waves by vortical
motions (Ford, personal communication, 1996). However, the $\pa (
\psi , \nabla^2 \psi )$ term in (8.1) can be put into an
explicitly quadrupolar form, which leaves only the $\pa ( \psi ,
\eta )$ term as a dipolar source. For weak waves, $\eta$ is
approximately given by $b (1+B^2) \psi$ (this is the
quasi-geostrophic approximation), so this term may be expected to
vanish to leading order in $\epsilon$. Thus it may well be that
the gravity-wave source term in (8.1) has reasonable scaling
properties in $\epsilon$. Further investigation of this point,
through numerical simulations, would be of considerable interest.

Because the energy of our weak-wave model is quadratic, and the
potential vorticity is a linear function of the other variables,
this means that one can apply standard spectral arguments to
deduce statistical mechanical equilibria and turbulent spectral
regimes. In fact, Warn (1986) and Farge \& Sadourny (1989) have
already done so. Although they were studying the shallow-water
equations, they made the approximation of a quadratic energy and a
linear potential vorticity in order to facilitate their analysis.
It follows that their results apply directly to our weak-wave
model.

\bigskip
\bigskip\noindent
{\it Acknowledgements.} This paper is the result of research begun
at the 1993 Summer Study Program in Geophysical Fluid Dynamics at
the Woods Hole Oceanographic Institution. We would thus like to
thank Rick Salmon, the Director of the Program, for making this
collaboration possible. We are also most grateful to Rupert Ford
for many extremely helpful comments.
The extension of the Lighthill theory to rotating shallow water
was first performed by W. A. Norton (unpublished).
TGS acknowledges support from
the Natural Sciences and Engineering Research Council and the
Atmospheric Environment Service of Canada.

\bigskip\bigskip

\centerline{\bf Appendix. On the non-existence of Arnol'd-stable
circular}
\centerline{\bf vortices in unbounded shallow-water flow}
\bigskip
Using the Arnol'd-stability methodology followed in Section 5, Ripa
(1987) derived a stability condition for axisymmetric basic
flows in the context of the shallow-water equations. In terms of
dimensional variables, Ripa (1987) showed that the
small-amplitude (quadratic) approximation to the
combined energy/angular-momentum/Casimir invariant $\P$ was
positive definite if there existed a constant $\alpha$ such that
$$
(V + \alpha r) \, {d Q \over dr} > 0 \quad \forall r \qquad
\hbox{and} \qquad (V + \alpha r)^2 < gH \quad \forall r
.\eqno\hbox{(A.1a,b)}
$$
Here $V(r)$ is the basic-state azimuthal velocity, $Q(r) = (f +
dV/dr + V/r)/H$ is the basic-state potential vorticity, and $H(r)$
is the basic-state total height, satisfying the gradient-wind
balance condition $g\, dH/dr = fV + V^2/r$. Conditions (A.1a,b) are
clearly analogous to the stability conditions (5.8a,b) for our
weak-wave model, after the appropriate non-dimensionalization.

However, it turns out that for physically realizable flows
in an unbounded domain (A.1a,b), like (5.8a,b), can {\it never\/} be
satisfied. (Ford (1994$a$) has provided a heuristic argument to this
effect for the special case of monotonic potential-vorticity
profiles.) We confine our
attention, as in Section 5$b$, to localized vortices with $dV/dr \to
0$ as $r \to \infty$. In that case
$H(r) = o( r^2 )$ as $r \to\infty$, and so the only way that (A.1b)
can be satisfied in an unbounded domain is by taking $\alpha = 0$.
In that case the angular momentum invariant is not being used, and
as Ripa (1987) argues, Arnol'd stability therefore cannot be
proven. The argument involves an appeal to the result of Andrews
(1984): if $\P$ consists only of the energy and a Casimir
invariant, neither of which are altered by a spatial translation of
the vortex, then the basic state cannot be a true extremum of $\P$
for arbitrary perturbations, and $\P$ cannot be sign definite.

There are, however, some subtleties involved with the use of
Andrews' theorem in an unbounded domain (Carnevale \& Shepherd
1990), and it is therefore desirable to establish the
non-existence of Arnol'd-stable circular vortices by direct
methods, as was done in Section 5$b$ for the weak-wave model.
By definition, $rV(r) = \int_0^r (QH - f ) r'
\, dr'$; this implies that $V = O(r)$ as $r \to 0$ (for finite
vorticity) and that $dV/dr \to 0$ as $r \to\infty$ (for a
localized vortex, with $QH - f \to 0$ as $r \to\infty$). Taking
$\alpha = 0$ as argued above, we now try to show that (A.1a,b) can
never be satisfied. Noting that
$$
{dQ \over dr} = {1\over H} \Bigl( {d^2 V \over dr^2} + {1\over r}
\, {dV \over dr} - {V \over r^2} \Bigr) - {Q \over gH} \Bigl( fV +
{V^2 \over r} \Bigr) ,\eqno\hbox{(A.2)}
$$
we obtain
$$
\eqalignno{
\int_0^\infty HV \, {dQ \over dr} \, dr &= - \int_0^\infty \biggl\{
\Bigl( {dV \over dr} \Bigr)^2 + {V^2 \over 2r^2} + {1 \over gH} \,
\Bigl( f + {dV \over dr} + {V \over r} \Bigr) \Bigl( fV^2 + {V^3
\over r} \Bigr) \biggr\} \, dr \cr
&= - \int_0^\infty \biggl\{
\Bigl( 1 - {V^2 \over 4gH} \Bigr) \Bigl( {dV \over dr} \Bigr)^2 +
{V^2 \over 2r^2} + {V^2 \over gH} \, \Bigl( f + {1\over 2} \, {dV
\over dr} + {V \over r} \Bigr)^2
\biggr\} \, dr .\cr&{}&\hbox{(A.3)}
}
$$
Now, (A.1b) requires $V^2 < gH$, in which case the right-hand side
of (A.3) is negative definite; but (A.1a) requires the left-hand
side of (A.3) to be positive definite. It follows that
Ripa's stability conditions (A.1a,b) can never be satisfied in an
unbounded domain.

\vfill\eject
\hoffset=0.5truecm
\parindent=-0.5truecm
\centerline{\bf References}
\bigskip
Andrews, D.G. 1984 On the existence of nonzonal flows satisfying
sufficient conditions for stability. {\it Geophys.Astrophys.Fluid
Dyn.}, {\bf 28}, 243--256.

Arnol'd, V.I. 1966 On an a priori estimate in the theory
of hydrodynamical stability. {\it
Izv.Vyssh. Uchebn.Zaved.Matematika}, {\bf 54}, no.5, 3--5. (English
transl.: {\it Amer.Math.Soc.Transl., Series 2}, {\bf
79}, 267--269 (1969).)

Benjamin, T.B. 1984 Impulse, flow force and variational
principles. {\it IMA J.Appl.Math.}, {\bf 32}, 3--68.

Broadbent, E.G. \& Moore, D.W. 1979 Acoustic destabilization of
vortices. {\it Phil.Trans.Roy. Soc.Lond.A}, {\bf 290}, 353--371.

Carnevale, G.F. \& Shepherd, T.G. 1990  On the
interpretation of Andrews' theorem. {\it Geophys.Astrophys.Fluid
Dyn.}, {\bf 51}, 1--17.

Farge, M. \& Sadourny, R. 1989 Wave-vortex dynamics in rotating
shallow water. {\it J.Fluid Mech.}, {\bf 206}, 433--462.

Fj\o rtoft, R. 1950 Application of integral theorems in deriving
criteria of stability for laminar flows and for the baroclinic
circular vortex. {\it Geofys.Publ.}, {\bf 17}, no.6, 1--52.

Ford, R. 1993 Gravity wave generation by vortical flows in a
rotating frame. Ph.D. thesis, University of Cambridge, 289 pp.

Ford, R. 1994$a$ The instability of an axisymmetric vortex with
monotonic potential vorticity in rotating shallow water. {\it
J.Fluid Mech.}, {\bf 280}, 303--334.

Ford, R. 1994$b$ Gravity wave radiation from vortex trains in
rotating shallow water. {\it J.Fluid Mech.}, {\bf 281}, 81--118.

Ford, R., McIntyre, M. E. \& Norton, W. A. 1996
Balance and the slow quasi-manifold: some explicit results.
{\it J. Atmos. Sci.} (to be submitted).

Kelvin, Lord 1880 On the oscillations of a columnar vortex. {\it
Phil.Mag.}, {\bf 10}, 155--168.

Lorenz, E.N. 1960 Energy and numerical weather prediction. {\it
Tellus}, {\bf 12}, 364--373.

Lorenz, E.N. 1986 On the existence of a slow manifold. {\it
J.Atmos.Sci.}, {\bf 43}, 1547--1557.

Nore, C. 1994 A Hamiltonian weak-wave model for shallow water
flow. In {\it Geometrical methods in fluid dynamics} (R. Salmon \&
B. Ewing-Deremer, eds.), Woods Hole Oceanog. Inst. Tech. Report
WHOI-94-12, pp.224--238.

Pedlosky, J. 1987 {\it Geophysical Fluid Dynamics}, 2nd ed.
Springer-Verlag, 710 pp.

Ripa, P. 1983 General stability conditions for zonal flows in a
one-layer model on the $\beta$-plane or the sphere. {\it J.Fluid
Mech.}, {\bf 126}, 463--489.

Ripa, P. 1987 On the stability of elliptical vortex solutions of
the shallow-water equations. {\it J.Fluid Mech.}, {\bf 183},
343--363.

Sadourny, R. 1975 The dynamics of finite-difference models of the
shallow water equations. {\it J.Atmos.Sci.}, {\bf 32}, 680--689.

Salmon, R. 1983 Practical use of Hamilton's principle. {\it
J.Fluid Mech.}, {\bf 132}, 431--444.

Shepherd, T.G.  1987  Non-ergodicity of inviscid
two-dimensional flow on a beta-plane and on the surface of a
rotating sphere. {\it J.Fluid Mech.}, {\bf 184}, 289--302.

Shepherd, T.G. 1988 Rigorous bounds on the nonlinear
saturation of instabilities to parallel shear flows. {\it J.Fluid
Mech.}, {\bf 196}, 291--322.

Shepherd, T.G. 1990 Symmetries, conservation laws, and
Hamiltonian structure in geophysical fluid dynamics. {\it
Adv.Geophys.}, {\bf 32}, 287--338.

Shepherd, T.G. 1992 Arnol'd stability applied to fluid
flow: successes and failures. In {\it Nonlinear Phenomena in
Atmospheric and Oceanic Sciences} (G.F. Carnevale \& R.T.
Pierrehumbert, eds.), 187--206. Springer-Verlag.

Shepherd, T.G. 1994 Applications of Hamiltonian theory to
geophysical fluid dynamics. In {\it Geometrical methods in fluid
dynamics} (R. Salmon \& B. Ewing-Deremer, eds.), Woods Hole
Oceanog. Inst. Tech. Report WHOI-94-12, pp.113--152.

Warn, T. 1986 Statistical mechanical equilibria of the shallow
water equations. {\it Tellus}, {\bf 38A}, 1--11.

Warn, T., Bokhove, O., Shepherd, T.G. \& Vallis, G.K. 1995
Rossby-number expansions, slaving principles, and balance
dynamics. {\it Quart.J.Roy.Meteor.Soc.}, {\bf 121}, 723--739.

Yuan, L. \& Hamilton, K. 1994 Equilibrium dynamics in a
forced-dissipative $f$-plane shallow-water system. {\it J.Fluid
Mech.}, {\bf 280}, 369--394.

\bye